\documentclass[aip,preprint,nofootinbib]{revtex4-2}
\usepackage{bm}
 \usepackage{tikz}
 \usetikzlibrary{arrows.meta}
 \usetikzlibrary{calc}
\usepackage{amsmath,amssymb}
\usepackage{xcolor}
\usetikzlibrary{decorations.markings}
\usepackage[colorlinks=true,linkcolor=blue,citecolor=blue,urlcolor=blue]{hyperref}
\usetikzlibrary{arrows.meta,calc,decorations.markings}
\tikzset{
  current/.style={
    postaction={decorate},
    decoration={markings,
      mark=at position #1 with {\arrow{Latex}}
    }
  }
}
\begin{document}

\title{Self-Force of a Dirac String: An Explicit Calculation}
\author{Alberto G. Rojo}
\affiliation{ Department of Physics, Oakland University, Rochester, MI 48309}

\maketitle

\noindent

\maketitle

\noindent

The Comment by McDonald~\cite{McDonald2026} shows how the  force between two strings derived in Ref.~\onlinecite{Rojo2026} can  be used to illuminate a subtle limitation of the Dirac monopole model.  Dirac~\cite{Dirac1948} already noted that a string would experience a nonvanishing and, in fact, divergent self-force.  At first sight this conclusion may seem surprising, since the Dirac string closely resembles an ordinary solenoid, whose self-force vanishes.  McDonald's argument is concise and insightful, but it relies on a somewhat sophisticated application of the two-string result.  In the same spirit as Ref.~\onlinecite{Rojo2026}, which presented an elementary derivation of the monopole–monopole interaction, we offer here an elementary derivation of a string's self-force.

We model a Dirac string as a semi-infinite solenoid of radius $a$, carrying a
current $I$ with $n$ turns per unit length, so that the magnetic flux inside
is $\Phi = \mu_0 n I \pi a^2$. 
%I added the following sentence.
(In the monopole limit, the magnetic charge $g = \Phi / \mu_0$ is a constant while $a \rightarrow 0$ and $I \rightarrow \infty$.) 
The self-force along the axis ($z$-direction) arises from the interaction of each current loop with the radial
magnetic field, $B_\rho$, produced by all other loops in the solenoid.

The radial component of the magnetic field at a point on the surface
of the solenoid can be extracted from the azimuthal vector potential via
\begin{equation}
    B_\rho(a, z) = -\frac{\partial A_\phi}{\partial z}. %I changed the coordinate point to a
    \label{eq:Brho}
\end{equation}
We will show that, at any $z$, this derivative reduces to the value of the vector potential due to the differentially small loop at the end of the solenoid $(z = 0)$.

The vector potential can be found from an integral over the current $\mathbf I = I \hat{\boldsymbol{\phi}}$: 
%Since I is a constant, it doesn't need to be a function of position.
\begin{eqnarray}
\mathbf A(a,z) &=& \frac{\mu_0}{4 \pi} \int_0^{\infty} n\,dz' \int_0^{2 \pi} \frac {I \hat{\boldsymbol{\phi'}}}{| \mathbf{r} - \mathbf{r'} |} a\,d \phi' \nonumber \\
&\equiv& \frac{\mu_0 na}{4 \pi} \int_0^{\infty} \,dz' \int_0^{2 \pi} \frac {I \hat{\boldsymbol{\phi'}} d \phi' }{\sqrt{2a^2(1-\cos\phi')+(z-z')^2}}   \nonumber
\end{eqnarray}
By symmetry, this result will be purely in the azimuthal direction. 

Because the solenoid is semi-infinite, as shown in Fig. 1, the difference between $\mathbf A(a,z)$ and $\mathbf A(a,z+dz)$ is simply:
\begin{equation}
\mathbf A(a,z+dz) - \mathbf A(a,z) =  dz \frac{\mu_0 n a}{4 \pi} \int_0^{2 \pi} \frac {I \hat{\boldsymbol{\phi'}} d \phi' }{{\sqrt{4a^2\sin^2(\phi'/2)+z^2}}}  
%I reformatted the fraction in front of the integral sign. But I'm confused by the negative sign.  Can you explain it?
%There may be a cancelling sign change in Eq. 4, since with I along \phi and an outward radial B, F would be along -z  
\end{equation}
and therefore $B_\rho(a,z) = -\partial A_{\phi} / \partial z$ is equal to the azimuthal $A_\phi$ due to this bottom loop:

\begin{equation}
    B_\rho(z) 
    = -\frac{\mu_0 n I a}{4\pi}
    \int_0^{2\pi}
    \frac{\cos\phi'}{\sqrt{4a^2\sin^2(\phi'/2) + z^2}}\, d\phi'.
    \label{eq:Bsol}
\end{equation}

The total self-force along the axis is obtained by integrating the axial force
on each current loop. A single loop of radius $a$ carrying current $I \hat{\boldsymbol{\phi}}$ in a
radial field $B_\rho  \hat{\boldsymbol{\rho}}$ experiences an axial force $-2\pi a I B_\rho  \hat{\mathbf{k}}$, so
integrating over all loops gives
\begin{eqnarray}
    F_z &=& -2\pi a\, n I \int_0^\infty B^{\text{sol}}_\rho(z)\, dz \\
    &=& \frac{\mu_0 n^2 I^2 a^2}{2}
    \underbrace{
        \int_0^\infty dz \int_0^{2\pi}
        \frac{\cos\phi'}{\sqrt{4a^2\sin^2(\phi'/2) + z^2}}
    }_{\displaystyle = \,\pi}.
    \label{eq:Fz}
\end{eqnarray} In the Supplementary material we show that the double integral equals $\pi$. 

Expressing this in terms of the magnetic flux $\Phi = \mu_0 n I \pi a^2$,
we obtain the compact result
\begin{equation}
    {F_z = \frac{\Phi^2}{2\pi\mu_0 a^2}.}
    \label{eq:selfforce}
\end{equation}
In the Dirac string limit $a \to 0$ with $\Phi$ held fixed, the self-force
diverges as $1/a^2$, confirming the observation of McDonald~\cite{McDonald2026}
and the concern raised by Dirac himself~\cite{Dirac1948}.
Equation~\eqref{eq:selfforce} makes the nature of this divergence explicit:
it is the unavoidable price of concentrating a finite magnetic flux into a
vanishing cross-section.
% ──────────────────────────────────────────────────────────────────────────────
It may seem surprising that the semi-infinite solenoid considered here experiences a nonvanishing self-force even before taking the Dirac string limit $a\to 0$.  A finite solenoid, of course, experiences no net self-force, since the magnetic stresses at its two ends cancel.  The nonzero result obtained here reflects the fact that a semi-infinite solenoid is an idealization: by removing one end, one also removes the compensating magnetic stresses required for overall momentum balance\footnote{The Maxwell stress tensor, applied to the semi-infinite solenoid, enables a more compact derivation of the self-force for more advanced students; this is presented in the supplementary material.
}.  In that sense, the semi-infinite configuration should be viewed as a limiting construction rather than as a physically realizable object.  
In summary, the self-force of a Dirac string can be understood as the edge
force of a solenoid once the cancellation between its two ends is removed.
Holding the magnetic flux fixed while shrinking the cross section makes the
associated magnetic pressure diverge, leading to the $1/a^2$ behavior obtained above.

% ─────────────────────────────────────────────────────────────────────────────
\section*{Acknowledgements}

The author thanks David Garfinkle for useful conversations.

% ─────────────────────────────────────────────────────────────────────────────

%%% ─────────────────────────────────────────────────────────────────────────────

\begin{figure}[t]
\centering

\begin{tikzpicture}[
    scale=1,
    line cap=round,
    line join=round,
    >=Latex,
    current/.style={
        postaction={decorate},
        decoration={
            markings,
            mark=at position 0.65 with {\arrow{Latex}}
        }
    }
]

% PARAMETERS (plain numbers)
\pgfmathsetmacro{\RR}{1.2}      % loop radius
\pgfmathsetmacro{\RYY}{0.45}    % squash (perspective)
\pgfmathsetmacro{\pitch}{0.4}  % spacing
\def\N{15}                      % number of loops

% AXES
\draw[->] (-1.6,0) -- (2.6,0) node[right] {$y$};
%\draw[->] (0,0) -- (0,6.2); 
%node[above] {$z$};
\draw[->] (0,0) -- (-1.6,-1.2) node[below] {$x$};

% CYLINDER AS STACK OF LOOPS
\foreach \k in {0,...,\N}
{
    \pgfmathsetmacro{\zz}{\k*\pitch}

    % darker loops near the end
    \ifnum\k<1
        \draw[current,line width=1.7pt] (0,\zz) ellipse ({\RR} and {\RYY});
    \else
        \draw[current,line width=0.6pt] (0,\zz) ellipse ({\RR} and {\RYY});
    \fi
}

% central axis
\draw[->,line width=0.7pt] (0,-0.2) -- (0,\N*\pitch+1.) node [right] {$z$};

% end marker
\node[left] at (-1.75,0) {$z=0$};

% indicate continuation (semi-infinite)
\node at (0,\N*\pitch+1.2) {$\vdots$};
\node at (0,\N*\pitch+1.6) {$\vdots$};

% bracket showing z
\draw[<->] (2.0,0.1) -- (2.0,3.7);
\node[right] at (2.0,1.9) {$z$};

\draw[thick,<-] (1.2,3.7)--(2.5,3.7) node [midway,above]{$B_\rho(a,z)$};
\end{tikzpicture}

%I added some text. I also added the parameter $a$ to specify the position.
{\caption{Semi-infinite cylinder represented as a stack of current loops. 
Arrows indicate the direction of the circulating current of magnitude $I$. 
The integrals to find $A(a,z)$ and $A(a,z+dz)$ are nearly identical; they 
differ only due to the inclusion in the second integral of the ring located 
in the range $z = 0 \rightarrow dz$. 
Since $B_\rho=(\nabla\times\mathbf{A})_\rho=-\partial A_\phi/\partial z$, 
the magnitude of the radial component of the magnetic field at height $z$ 
on the surface of the semi-infinite cylinder, $B_\rho(a,z)$, is equal to the 
magnitude of the vector potential $A_\phi(a,z)$ produced by the ring at the 
bottom of the cylinder.}}
\label{fig:loopstack}
\end{figure}
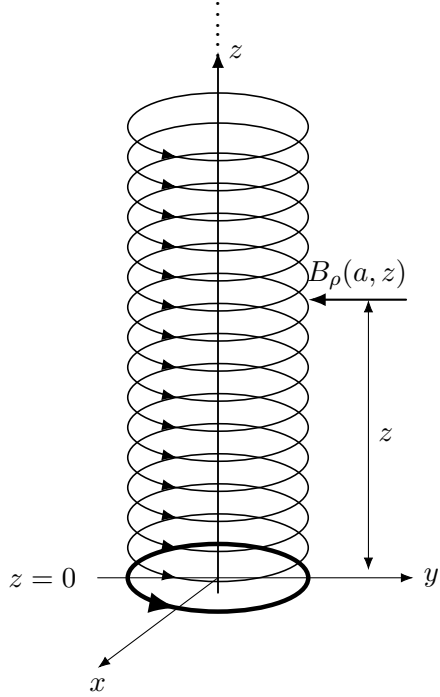


\begin{thebibliography}{9}

\bibitem{McDonald2026}
K.~T.~McDonald, ``Self force on Dirac monopoles,''
Am.\ J.\ Phys.\ (to be published, 2026).

\bibitem{Rojo2026}
A.~G.~Rojo, ``Coulomb force between two Dirac monopoles,''
Am.\ J.\ Phys.\ \textbf{94}, 65--67 (2026).
\href{https://doi.org/10.1119/5.0244845}{https://doi.org/10.1119/5.0244845}

\bibitem{Dirac1948}
P.~A.~M.~Dirac, ``The theory of magnetic poles,''
Phys.\ Rev.\ \textbf{74}, 817--830 (1948).
\href{https://doi.org/10.1103/PhysRev.74.817}{https://doi.org/10.1103/PhysRev.74.817}

\bibitem{GR2015}
I.~S.~Gradshteyn and I.~M.~Ryzhik,
\textit{Table of Integrals, Series, and Products}, 8th ed.\
(Academic Press, 2015), formula 1.441.2.

\end{thebibliography}
\end{document}